# Effects of model approximations for electron, hole, and photon transport in swift heavy ion tracks

R.A. Rymzhanov[1], N.A. Medvedev[2,3*], A.E. Volkov[1,4,5,6,7]

[1] *Joint Institute for Nuclear Research, Joliot-Curie 6, 141980 Dubna, Moscow Region, Russia*

[2] *Department of Radiation and Chemical Physics, Institute of Physics, Czech Academy of Sciences, Na Slovance 2, 182 21 Prague 8, Czech Republic*

[3] *Laser Plasma Department, Institute of Plasma Physics, Czech Academy of Sciences, Za Slovankou 3, 182 00 Prague 8, Czech Republic*

[4] *National Research Centre 'Kurchatov Institute', Kurchatov Sq. 1, 123182 Moscow, Russia*

[5] *Lebedev Physical Institute of the Russian Academy of Sciences, Leninskij pr., 53,119991 Moscow, Russia*

[6] *National University of Science and Technology MISiS, Leninskij pr., 4,119049 Moscow, Russia*

[7] *National Research Nuclear University MEPhI, Kashirskoye sh., 31, 115409 Moscow, Russia*

## Abstract

The event-by-event Monte Carlo code, TREKIS, was recently developed to describe excitation of the electron subsystems of solids in the nanometric vicinity of a trajectory of a nonrelativistic swift heavy ion (SHI) decelerated in the electronic stopping regime. The complex dielectric function (CDF) formalism was applied in the used cross sections to account for collective response of a matter to excitation. Using this model we investigate effects of the basic assumptions on the modeled kinetics of the electronic subsystem which ultimately determine parameters of an excited material in an SHI track.

In particular, (a) effects of different momentum dependencies of the CDF on scattering of projectiles on the electron subsystem are investigated. The 'effective one-band' approximation for target electrons produces good coincidence of the calculated electron mean free paths with those obtained in experiments in metals. (b) Effects of collective response of a lattice appeared to dominate in randomization of electron motion. We study how sensitive these effects are to the target temperature. We also compare results of applications of different model forms of (quasi-) elastic cross sections in simulations of the ion track kinetics, e.g. those calculated taking into account optical phonons in the CDF form vs. Mott's atomic cross sections. (c) It is demonstrated that the kinetics of valence holes significantly affects redistribution of the excess electronic energy in the vicinity of an SHI trajectory as well as its conversion into lattice excitation in dielectrics and semiconductors. (d) It is also shown that induced transport of photons originated from radiative decay of core holes brings the excess energy faster and farther away from the track core, however, the amount of this energy is relatively small.

Keywords: swift heavy ion, electronic stopping, TREKIS, Monte Carlo, electronic kinetics, photon transport

---

* Corresponding author. Email: nikita.medvedev@fzu.cz

# I. Introduction

In this paper we analyze model assumptions used in the Monte-Carlo (MC) codes describing electron kinetics in highly excited systems. We perform our analysis on example of the recently developed code modeling Time Resolved Electron Kinetics in swift-heavy-ion Irradiated Solids (TREKIS) [1]. The model describes excitation of a solid by a nonrelativistic swift heavy ion (SHI) decelerated in the electronic stopping regime as well as further spreading of generated electrons and holes and their interaction with matter in the vicinity of the ion trajectory.

The present work extends the original model [1] by a detailed analysis of solid state effects on scattering processes of appearing electrons, valence holes, and photons generated in an ion track by decay of core holes. Focusing on the energies of swift heavy ions near the Bragg peak, we investigate how these processes affect temporal and spatial evolution of the electron ensemble and its interaction with the atomic subsystem of a material.

The MC method uses cross sections of interactions of a projectile with an ensemble of scattering centers. Due to spatial and temporal correlations, these scattering cross sections depend on collective response of the electronic and atomic subsystems of a solid target to perturbations caused by a projectile.

Cross sections of interaction of a projectile with a solid can be calculated precisely from the first-principle models that take into account the band structure of a target and other collective effects [2–4]. But such techniques require great computational resources and are difficult to combine with other codes. Thus, approximate methods prevail so far in calculations of charged particle cross sections in solids [5–9]. An analysis of effects of different model forms of the cross sections (or the mean free paths, MFP) of inelastic and elastic scattering of electrons and holes, evaluated by different methods, is necessary to select the most important channels of the relaxation kinetics of the extremely excited electron subsystem in SHI tracks.

Also, a study of the cross sections of interaction of a particle with a solid has a practical interest, because they are widely used for interpretation of experiments dealing with propagation of electrons in materials, e.g. Auger electron spectroscopy [10], X-ray photoemission spectroscopy [11], low-energy electron diffraction and Bremsstrahlung isochromat spectroscopy [9], etc.

In particular, within the first Born approximation (the first order of the perturbation theory of the kinetic energy of a projectile), effects of collective response of a condensed target can be taken into consideration expressing the scattering cross sections in terms of the Dynamic Structure Factor (DSF) [12]. The fluctuation-dissipation theorem links the DSF to the Complex Dielectric Function (CDF) of a material [13–15].

The CDF of a material can be calculated using theoretical or semi-empirical models [4,16,17]. One of the well-known approaches was developed in [9,18,19] using the Lindhard-type dielectric function within the random phase approximation, and was later replaced by Mermin dielectric function [20], extended further by the so-called full-conserving dielectric function [21].

It is important that the dependencies of the CDF on the transferred energy can be obtained from the experimental optical data. Ritchie and Howie [8] demonstrated that these data can be very well fitted when CDF is approximated by a set of artificial Drude-like oscillators. Such CDFs are often used for calculations of electron mean free paths (MFPs) [22,23], as well as in many other applications [5,6,8,9], in particular for quantitative evaluation of various effects in electron spectroscopy [24].

The CDF from the optical data provides appropriate cross sections but, unfortunately, does not contain the dependence on the momentum transferred during a collision, because in the optical limit the wave vector transferred to the scattering system is equal to zero ($q$ = 0) within the dipole approximation. We use additional assumptions about the dispersion dependencies of the artificial oscillators [25] ($q(E)$, see below) to model the dependence on momentum of the CDF fitted from the optical data: free-particle, single-pole approximation [18], momentum dependence proposed by Ritchie *et al.* [8], and the effective one-band approximation [26].

This forms the problem of an accuracy of the introduced approximations to be studied. We investigate this effect of application of different dispersion dependencies of the artificial oscillators used for fitting of CDF obtained in optical experiments. We compare the calculated mean free paths of electrons calculated under different assumptions about the dispersion law $q(E)$ in different materials with those from experiments and other simulations. We also investigate effect of the model forms of the dispersion laws in the applied CDF on transient spatial distributions of electrons and valence holes densities in SHI tracks.

The elastic scattering of electrons and valence holes generated in a track governs transfer of the excess electronic energy into the lattice resulting in its excitation [27]. Throughout this paper, the term 'elastic scattering' refers to scattering events with exchange of the kinetic energy only, without transfer into potential energy via inducing excitation or ionization events. Sometimes in the literature such processes are referred to as 'quasi-elastic' to emphasize that the energy of the incident particle is changing in the course of collision, albeit this energy change is small [28–30]. Such loss of energy in 'elastic' scattering events is taken into account in TREKIS [1].

Further relaxation of the excess lattice energy can cause transformations of the material structure in a track. Causing large changes of the momentum of an incident particle, this kind of electron and valence hole scattering thereby considerably influences their transport. This effect was demonstrated in experiments where LiF crystals were irradiated with gold ions at cryogenic and room temperatures [31,32]. A lack of phonons participating in scattering events at cryogenic lattice temperatures results in twice as large diffusion ranges of valence holes before their self-trapping.

In this paper we investigate effects of two limit forms of the elastic cross sections on electrons-to-lattice energy transfer rate, and spatial spreading of electronic excitations: (a) the cross sections corresponding to the optical phonons branch of the CDF restored from the experimental photo-absorption data [33], and (b) Mott's atomic cross sections with modified Molier's screening parameter [34].

Already within tens of femtoseconds after the projectile passage, appearing valence holes accumulate noticeable part of the excess electronic energy generated in a track [35,36]. We demonstrated that the kinetics of valence holes significantly affects redistribution of this energy in the vicinity of an SHI trajectory, as well as its conversion into a lattice excitation in dielectrics and semiconductors [36]. It was shown that valence holes provide the lattice with even more energy than generated free electrons in the track core, thus affecting strongly the kinetics of structure changes in an SHI track [36].

An effect of radiative decays of core holes in SHI tracks is also studied. Core-hole relaxation is dominated by the Auger decay for the excitation regimes studied here; nevertheless radiative decays are possible and thus their influence on the energy transport in tracks can affect the track kinetics. Photoemission by radiative decays is incorporated into TREKIS code [1] to analyse the effect of photon transport in SHI tracks. It is demonstrated quantitatively that even though photons can bring energy far from the track core, this effect is small due to small probabilities of radiative decays for the analyzed range of SHI energies.

## II.    Scattering cross sections

Within the assumed first Born approximation, when the kinetic energy of a projectile is much larger than the potential energy of its interaction [37], a cross section of scattering of an incident particle on a spatially and dynamically coupled system of scattering centers can be split into a product of the cross section of scattering on a single (isolated) scattering center and the Dynamic Structure Factor (DSF) [12]. DSF describes the collective response of a target to excitation. For scattering of charged particles, the DSF can be expressed in terms of the loss function of a target (imaginary part of the Complex Dielectric Function (CDF), $\varepsilon(\omega,q)$) via the fluctuation-dissipation theorem [13–15]. This results in the following form of the differential cross section, $\sigma$, of a charged particle interacting with a coupled system of charged scattering centers in the isotropic case assumed within MC models:

$$\frac{d^2\sigma}{d(\hbar\omega)d(\hbar q)} = \frac{2(Z_e(v,q)e)^2}{n_{sc}\pi\hbar^2 v^2}\frac{1}{\hbar q}\left(1-e^{-\frac{\hbar\omega}{k_B T}}\right)^{-1}\mathrm{Im}\left[\frac{-1}{\varepsilon(\omega,q)}\right], \quad (1)$$

where $e$ is the electron charge; $\hbar\omega$ is the transferred energy and $\hbar$ is the Planck's constant; $k_B$ is the Boltzmann constant, and $T$ is the temperature of the sample; $n_{sc}$ is the density of scattering centers.

$Z_e(v,q)$ is the effective charge of the projectile penetrating through a solid as a function of the ion velocity, $v$, and a transferred momentum, $q$. For an incident electron and a valence hole $Z_e = 1$. For an SHI we use a semi-empirical Barkas' formula for an effective ion charge, see the discussion about $Z_e$ in Ref. [1]. Application of this formula is based on the fact that the charge equilibration depth of an SHI in a solid typically does not exceed ~100 nm [38,39], which is much shorter than the total penetration depths of swift heavy ions in solids (~10-100 μm), thus, in the bulk the equilibrium dependence of the ion charge on the ion velocity can be assumed. Possible charge fluctuations or nonequilibrium states [40] are disregarded in the presented model. It should be noted that an investigation of the transient electronic kinetics in an excited solid is the main aim of the present work, but not the description of an SHI itself. Therefore, we avoid the problem of the exact velocity dependence of the charge of a penetrating ion, by using Barkas' formula. Although Barkas' effective charge does not correspond to a realistic ion charge in a matter, it systematically produces very good coincidence of the ion energy losses within the first Born approximation (calculated using the CDF-based cross sections with experimental ones [1]), and therefore can be considered essentially as a good fitting parameter for DSF-CDF formalism applied in this paper.

The factor $\left(1-e^{-\frac{\hbar\omega}{k_B T}}\right)^{-1}$ in Eq. (1) determines the dependence of the cross section on the temperature of the system of scattering centers [41–44]. At transferred energies $\hbar\omega \gg k_B T$ or at $T = 0$ K, Eq. (1) turns to the common expression of the differential cross section used in Ref. [1] (note that in the original paper [1] the density of scattering centers was omitted):

$$\frac{d^2\sigma}{d(\hbar\omega)d(\hbar q)} = \frac{2(Z_e(v,q)e)^2}{n_{sc}\pi\hbar^2 v^2} \frac{1}{\hbar q} \mathrm{Im}\left[\frac{-1}{\varepsilon(\omega,q)}\right],$$

We reconstruct the CDF from optical data as described in detail in Ref. [1]. The inverse imaginary part of CDF (the loss function) can be fitted by a set of artificial optical oscillators (zero momentum transfer) [8]:

$$\mathrm{Im}\left[\frac{-1}{\varepsilon(\omega,q=0)}\right] = \sum_{i=1}^{N^{os}} \frac{A_i \gamma_i \hbar\omega}{[\hbar^2\omega^2 - E_{0i}^2(q=0)]^2 + (\gamma_i \hbar\omega)^2}, \quad (2)$$

here, $E_{0i}$ means the characteristic energy of the $i$-th oscillator, $A_i$ is the fraction of electrons with energy $E_{0i}$, and $\gamma_i$ is the $i$-th energy damping coefficient.

As mentioned above, the CDF from the optical data does not contain the dependence on the momentum transferred during a collision, because in the optical limit the wave vector transferred to the scattering system is equal to zero ($q = 0$) within the dipole approximation. At least two methods are used to introduce this dependence into the CDF. One of this method, applied in this paper, is based on assumptions about the dispersion dependencies of the energy of artificial oscillators used in Eq.(2) fitting the CDF ($E_{0i}(q)$) [25].

Ashley in [19] proposed another method of introducing the momentum dependence into CDF by supplying of the transferred energy with a dispersion law ($\omega \rightarrow \omega(q)$) instead of that for the energy of artificial optical oscillators, $E_{0i}$. A similar idea (with different realization) is used in the so-called Penn's algorithm [9]. These approaches [19] and [9] are fundamentally different from that used in this paper, and thus will not be analyzed in the current work.

There are a few most-often used models of the dispersion dependencies $E_{0i}(q)$ of the artificial oscillators used in the fitting Eq.(2).

1) Very often these dispersion dependencies are taken as that of a free particle: $E_{0i}(q) = E_{0i} + \hbar^2 q^2/(2M)$ ($M$ is the mass of a scattering centre, which is $m_e$ in case of inelastic scattering on the electron subsystem, i.e. ionization events or a plasmon excitations; $m_{at}$, the average mass of target atoms, in case of elastic scattering). This dependence was used throughout Ref. [1]. As it was mentioned in [8], in the case of an incident electron, such model should work well at high electron energies (>100 eV) and at the limit of zero transferred energy ($\omega \rightarrow 0$). For the intermediate energies (~1 eV – 100 eV), the inelastic mean free paths of an electron calculated with Eq.(2) may differ from the experimental data, and, unfortunately, it is hard to say *a-priori* how large this difference can be for a selected material.

Discrepancies can arise within such approximation due to the following reasons:

a) Interaction of slow electrons with a solid target results in excitation of valence or conduction band electrons which dispersion relation differs from the free-particle energy-momentum dependence.

b) The dispersion laws of optical phonons and plasmons, which peaks are present in the CDF taken from the optical data, differ from that of a free particle.

c) In the discussed model, target electrons are assumed to be uniformly distributed point-like particles at fixed positions (instantaneous scattering). In a real medium, the electrons have non zero velocities and certain momentum distribution that can affect the cross sections of slow incident particles [45]. This effect is known in the literature as the 'shell correction' and has typical influence on energy losses ~5% [46].

In this paper we are investigating only the effects of the different dispersion laws assuming that the uniform distribution of electrons and instantenous collisions can be applied to describe target electrons during scattering.

In addition to the free particle approximation, the following model dispersion relations for artificial oscillators used in the fitting of CDF from the optical data can be applied in order to extend the optical loss function over the whole space of the transferred momentum and energy.

2) The single-pole approximation for electrons is often used in the literature [18]:

$$E_{0i}^2(q) = E_{0i}^2 + \frac{1}{3}v_f^2(\hbar q)^2 + \left(\frac{(\hbar q)^2}{2m_e}\right)^2, \qquad (3)$$

where $v_f$ is the Fermi velocity of electrons in the solid.

3) Ritchie *et al.* in [8] suggested to incorporate the *q*-dependence of target electrons directly into the optical oscillator parameters $E_{0i}$ and $\gamma_i$ by their substitution with $E_{0i}(q)$ and $\gamma_i(q)$ in the following form:

$$E_{i0}(q) = \left[E_{i0}^p + \left(\frac{(\hbar q)^2}{2m_e}\right)^p\right]^{\frac{1}{p}} \text{ and } \gamma_i(q) = \left[\gamma_i^2 + \left(\frac{(\hbar q)^2}{2m_e}\right)^2\right]^{\frac{1}{2}}, \qquad (4)$$

where $p = \frac{2}{3}$ according to [8].

4) Additionally, the scheme based on the evaluation of the dispersion relation of the valence or conduction bands electrons from the density of states (DOS) of the material within the 'effective one-band' approximation is proposed in [26]. The DOS of a material is defined as

$$D(E) = \frac{s}{2\pi^2}q^2(E)\frac{dq}{dE}, \qquad (5)$$

which depends on the dispersion relation $q(E)$, and where $s$ determines the spin factor equal to two for the electron system. According to [26], in the 'effective one-band' model this expression can be solved analytically for the assumed effective isotropic momentum:

$$q(E) = \sqrt[3]{\frac{6\pi^2}{s}\int_0^E d\varepsilon \cdot D(\varepsilon)} \qquad (6)$$

resulting in an average isotropic dispersion relation between the momentum *q* and the energy *E*. We propose to use this relation (6) for the inverse dependence $E(q)$ in formula (2) substituting $E_{0i}(q) = E_{0i} + E(q)$, for a realistic DOS of a material obtained within an *ab-initio* model or from experiments.

As an example, Figure 1 presents a comparison between different energy-momentum dependencies resulted from an application of the above-mentioned assumptions for gold. The effective one-band approximation is analytically continued as a free-electron dispersion curve above the maximum available DOS energy of the valence band (~9.5 eV), Ref. [47]. Because the DOS of gold differs significantly from the DOS of a free-electron gas, the dispersion curve calculated from the DOS also differs from that of the free-electron as well as from the single-pole approximation.

The same cross sections are used for valence holes transport, with their own effective mass, see Ref. [36] for details.

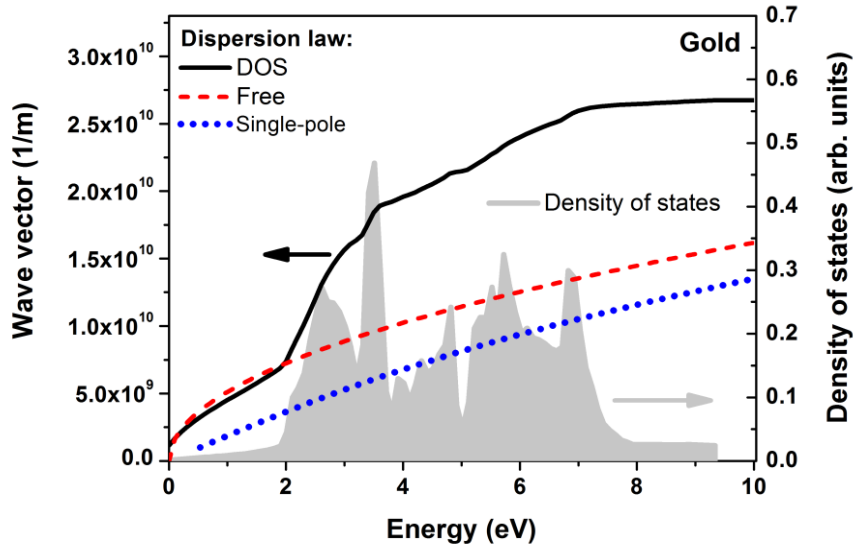

**Figure 1** The dispersion relations within different approximations and the density of states of gold (taken from Ref. [47]).

### III. Radiative decays of holes and photon transport

There are a number of channels of core holes decay: (a) intra-atomic Auger decay when an electron from an upper shell or the valence band fills the core hole, while another electron is released as a free electron carrying out the excess energy, which typically occurs within a few femtoseconds [48]; (b) Coster-Kronig decay, which is similar to Auger-decay with the difference that the hole is filled by an electron from a higher subshell of the same atomic shell. For deep shells of a heavy element Coster-Kronig decay can be faster than the Auger-decay [48]. (c) Inter-atomic Auger-decay, also known as Knotek-Feibelman transition [49], where the electrons involved in the transitions can be taken from a neighboring atom of a solid (in chemistry it is known as a class of processes named Interatomic Coulombic Decay, ICD, Excitation-transfer-ionization, ETI, and related processes [50,51]). (d) Radiative decay, or a fluorescence, in which a hole hops up to an upper shell emitting the excess energy as a photon [35,48].

The radiative decay is the most important channel for holes in deep shells of heavy elements, where the radiative life-time is shorter than that of Auger-decay [48]. Although for the SHI energies considered here, such deep shells are rarely ionized, it is important to consider such a possibility to investigate a contribution of generated photons and their transport into the SHI track creation process.

Tacking this into account, we added a possibility of radiative decay of core holes into TREKIS code in the following way. First, we choose which kind of decay a hole undergoes, in accordance with the life-times of different processes. An estimation of the life-time of radiative decay is obtained using Poisson's distribution – the same method as used for the Auger-decay case, with its own characteristic times taken from the EADL-database [52]. A shell which the hole is jumping to after the decay is chosen randomly among all the upper shells of all the atoms in the compound, or the valence band. Note also that, although EADL-database uses atomic data, for core shells they should be close to the case of solids, since deep atomic levels are only slightly affected by the collective effects.

The emitted energy is then propagating as a photon with the speed of light in vacuum. The photon mean free path until the photoabsorption is calculated from the total cross section. The cross sections of photoabsorption are taken from EPLD97-database [53] for each shell of each atom of the solid. In case of the valence band, the photoabsorption cross sections are evaluated from the complex dielectric function, Ref. [1] (inverse of Eq.(5) in there). The shell that absorbs the photon is chosen according to its relative photoabsorption cross section.

Note also that the plasmon peak of CDF indicates the cross section of photoabsorption by the valence or conduction band. For this case, similarly to the electron impact ionization [1], we assume that a produced plasmon decays instantly into a single electron-hole pair, thereby essentially resulting into an ionization event.

Presence of optical phonons in the CDF shows that a direct photoabsorption by collective atomic oscillation is possible in the material. This, however, occurs only for photons in the infrared range, which corresponds to the photon energies below ~0.1 eV. Photons with such energies could be produced in a semiconductor due to electron-hole recombination between the valence and the conduction band. However, as discussed above, this process requires very long timescales (typically longer than a nanosecond) in comparison to the relaxation time of the electronic subsystem in a track, and thus does not play a role in the considered cases. Since Bremsstrahlung emission is also neglected, no photons of the relevant energies are produced in SHI tracks, and thus photoabsorption by optical phonons can be neglected.

During photoabsorption a new hole is produced, and an electron is emitted with the energy equal to the difference between the ionization potential of the shell and the photon energy. The initial direction of motion of the electron is randomly chosen within the solid angle. All thereby produced electrons and holes are then modeled in exactly the same manner as all other primary and secondary electrons produced by an ion or electron/hole impact ionizations and Auger-decays.

## IV. Results and discussion

## 1. *Temperature dependence of electron mean free paths*

The complex dielectric function is sensitive to the band structure of the material and its statistics, and thus can also be changed by material heating. In the present version of TREKIS MC-code we use CDF functions restored from optical experiments made at the fixed (room) temperature. Therefore, in the estimations presented below we assumed that the CDF itself is not temperature dependent, and the whole effect of temperature comes solely from the pre-factor of the DSF, Eq.(1). In reality, the complex dielectric function is sensitive to the band structure of the material, and thus can also be changed by material heating. However, we expect such changes to be small [54], and thus such estimations are useful for the first investigations attempt of the temperature effect on the cross sections. At least inelastic scattering of electrons is affected relatively slightly by CDF changes caused by different lattice temperatures below melting, since even in case of phase transitions such changes appeared to be not too large [55].

An example of the calculated elastic mean free paths of electrons with respect to scattering on optical phonons, Eq.(1), in LiF for different temperatures of the lattice is presented in Figure 2. Here and further we use coefficients of the CDF from Ref. [1], where they were fitted for a number of different materials.

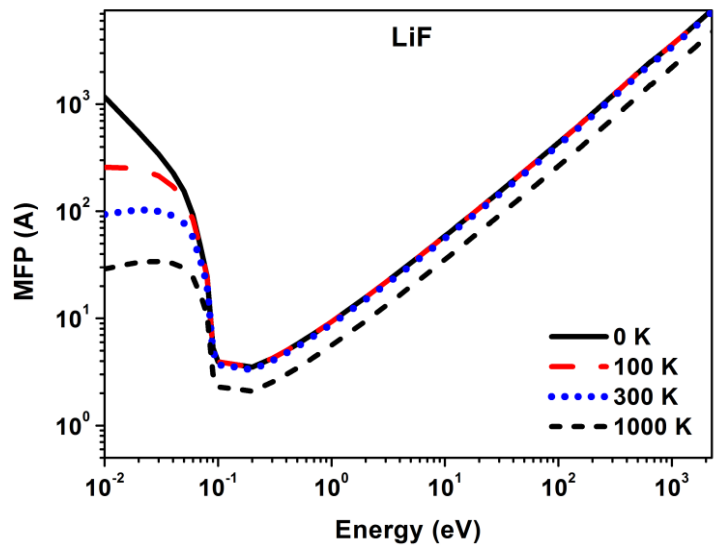

**Figure 2** The elastic mean free path (scattering on optical phonons only) of electrons in LiF at different temperatures of the target lattice.

The effect is most pronounced at low electron energies ($E_e<0.1$ eV). It should also be noted that the two-temperature thermal spike model [56–58] assumes interaction of electrons having such energies with phonons as the dominant mechanism of energy and momentum transfer from the excited electronic subsystem to the target lattice. This model is widely used by the society for description of damage effects in materials irradiated with swift heavy ions. The detected temperature dependence affects considerably the cross sections of the low-energy-electron coupling with a lattice which is the key parameter of this model.

Figure 3 illustrates the effect of the lattice temperature on the kinetics of excited electron subsystem in LiF irradiated with 2187 MeV Au ions. It demonstrates only minor influence of the temperature dependence of the prefactor in Eq.(1) on the electron kinetics even at temperatures close to the melting point of LiF.

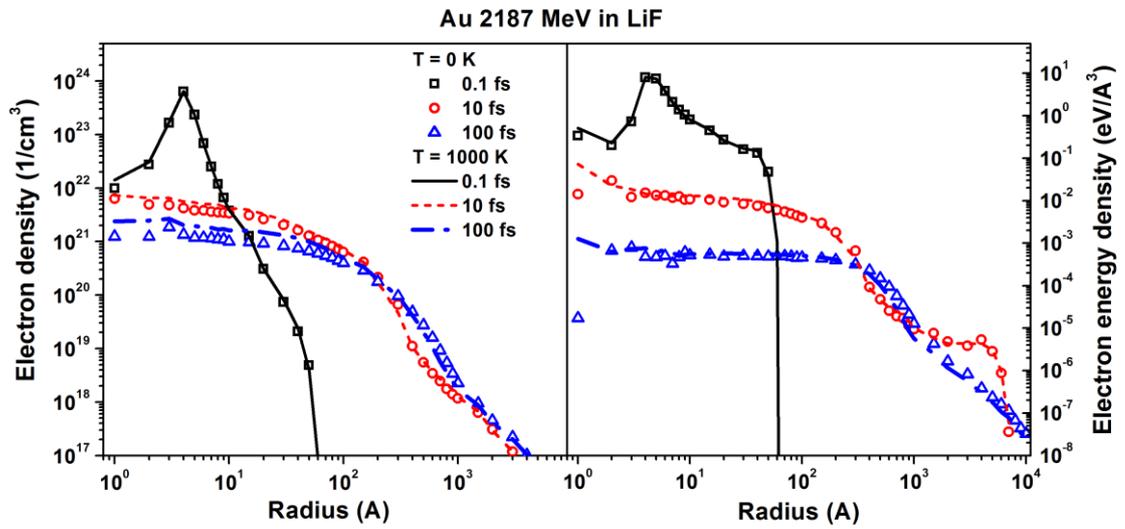

**Figure 3** The density of excited electrons and their energy at different times after passage of Au 2187 MeV ion in LiF calculated at T = 0 K and T = 1000 K target temperatures.

Inelastic cross sections of electrons in LiF are not affected by this temperature factor because the lower limit of transferred energy in inelastic collisions is much higher than the considered temperatures ($E_{gap}$ = 14.6 eV for LiF). But, in case of metallic targets, the low-energy part of the inelastic scattering of an incident electron on conduction band electrons may be sensitive to the electronic temperatures.

The inelastic mean free paths of electrons in gold at different electronic temperatures are presented in Figure 4. Equilibrium heating of electrons together with the lattice (up to 1000 K) does not produce noticeable changes in the electron inelastic mean free paths. In contrast, nonequilibrium heating of the electronic subsystem realized in e.g. laser spots or in an SHI track, elevates the electron temperatures to values much higher than those of the lattice temperature starting to affect electron inelastic MFPs of electrons.

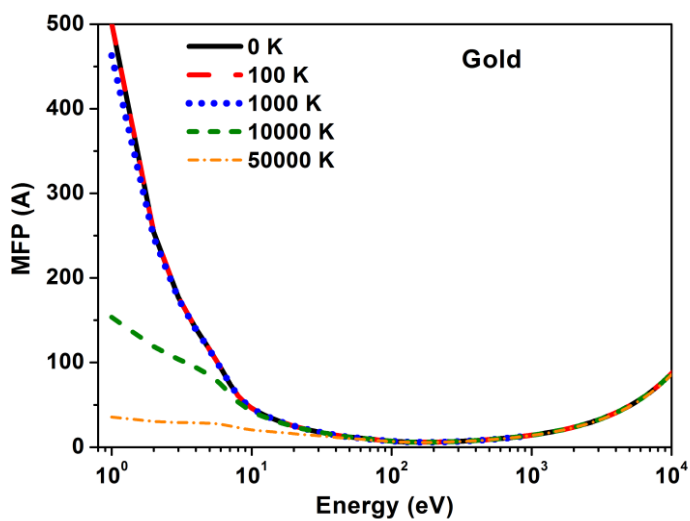

**Figure 4** The inelastic mean free path of electrons in gold at different temperatures of the electron subsystem of the target.

Figure 5 demonstrates that changes in the electron mean free paths in gold at high electronic temperature (50000 K) results in increase of the number of excited electrons in a track, but the energy density of the electronic ensemble is affected only slightly.

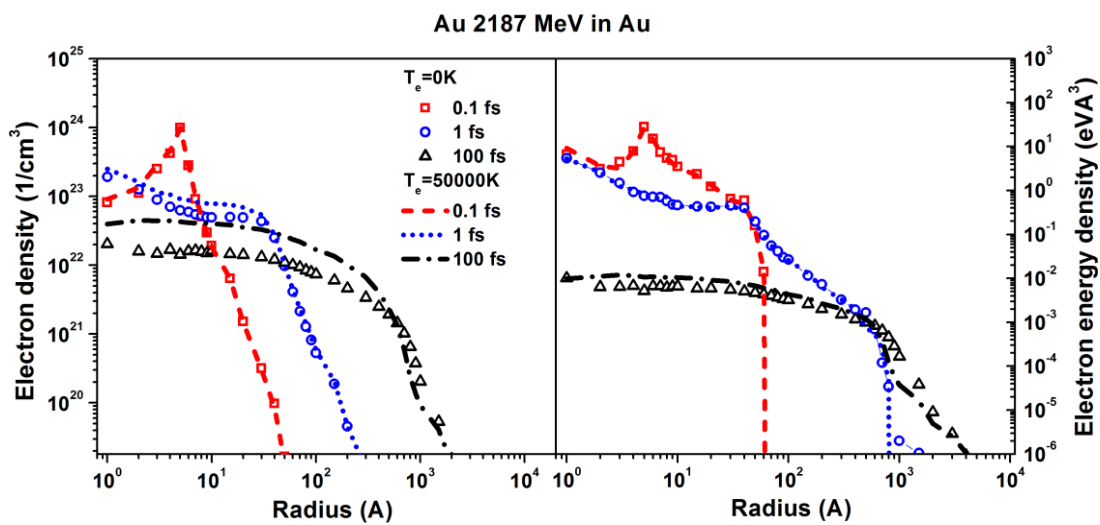

**Figure 5** The density of excited electrons and their energy at different times after passage of Au 2187 MeV ion in gold calculated at $T_e = 0$ K and $T_e = 50000$ K target electronic temperatures.

Considering that the effects of temperatures seem to be minor for the electronic kinetics in SHI tracks at femtosecond timescales, further in this work we are assuming target temperature $T = 0$ K for both, atomic and electronic ensembles of a target, in all shown calculations.

## 2. *Effect of momentum dependence on inelastic mean free path*

Figure 6 and Figure 7 demonstrate the inelastic mean free paths (IMFP) of electrons calculated taking into account different model forms of the dispersion relation of the energies of the artificial oscillators (Eqs.(3-6)) in the used cross sections describing electron scattering in noble metals: gold and copper. The calculated MFPs are compared to the experimental data, and to the free-electron approximation used in our previous work [1]. The comparison of the discussed dispersion models does not reveal any significant differences between the results of their applications for high electron energies. The low- and intermediate- energy parts of IMFP are affected more significantly. The model dependence proposed by Ritchie *et al.* [8] improves the coincidence only for low energy in gold (Figure 6), while for intermediate energies the agreement is getting worse (between ~5 eV and ~500 eV). The single-pole approximation works better for intermediate energies, but underestimates the IMFPs at low energies (< 5 eV). In contrast, our proposed approach based on the effective one-band model provides good correction to the IMFPs of an electron in noble metals, which coincide reasonably well with the experimental and calculated data in the entire range of electron energies.

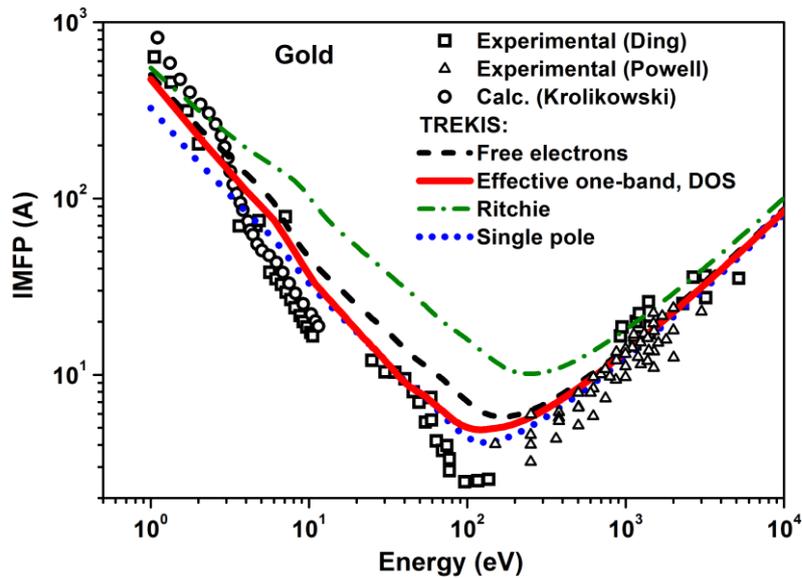

**Figure 6** The mean free paths of an electron in gold calculated with TREKIS code using different dispersion laws. The compilations of the experimental data were taken from the papers of Ding et al. [18] (open squares) and Powell et al. [6] (open triangles). The calculated data (circles) are from Krolikowski [59].

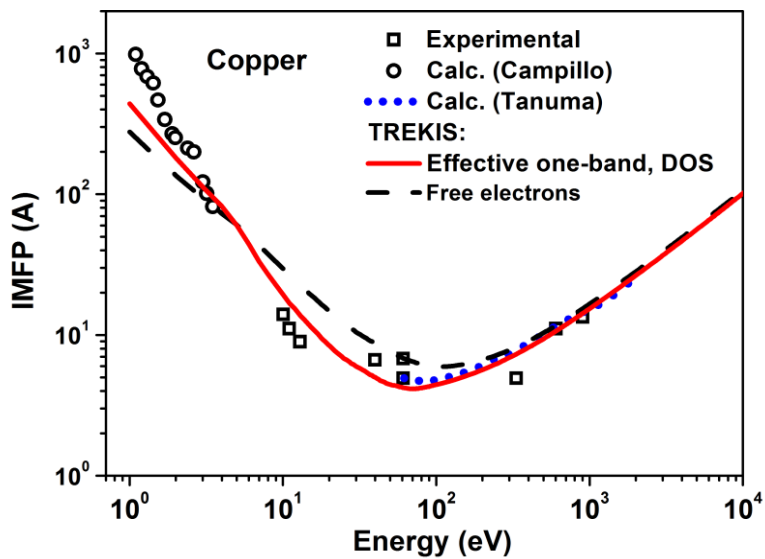

**Figure 7** The electron mean free paths in copper calculated with TREKIS code using different dispersion models. The experimental data (open squares) were taken from [9,60]. The calculated data from Campillo *et al.* (circles, Ref. [3]), and Tanuma *et al.* (blue dots, Ref. [5]), are shown for comparison.

Figure 8 demonstrates that in silicon, changing the momentum dependence of the CDF does not produce a better match of the calculated mean free paths of electrons with *ab-initio* calculations from [61] that took into account realistic band structure of a material. Unfortunately, only a few experimental data at low energies are available and no data for the intermediate electron energies exist to the best of our knowledge.

It seems that neither the free-particle approximation of the dispersion law for the oscillators applied in the fitted form of CDF, nor its improvements, can quantitatively reproduce the experimental electron MFPs in nonmetallic materials for low energy electrons. Only qualitative agreement is achieved at low energies of scattering electrons (below ~50 eV). Perhaps, this occurs because for such energies of an incident electron its own dispersion relation can differ from that of a free electron used here tending to that of the conduction band. Thus, not only the dispersion relation for target electrons, but for the incident electron too, should be modified. This is a matter for a future investigation.

However, such low energy electrons perform only a few ionization events until their energies drops below the ionization threshold. Thus, they are not expected to affect considerably the electron kinetics in a track. Therefore, in the current work we only use the models presented.

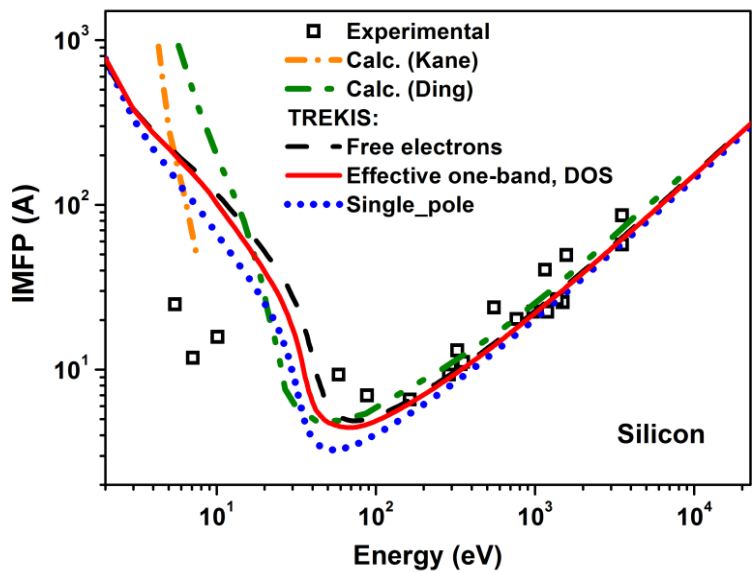

**Figure 8** The electron mean free paths in silicon calculated with TREKIS code using different dispersion models. Compilation of the experimental data (open squares) and calculations of Ding (green dash-dot-dotted line) were taken from [18]. *Ab-initio* calculation (Kane, orange dash-dotted line) is taken from [61].

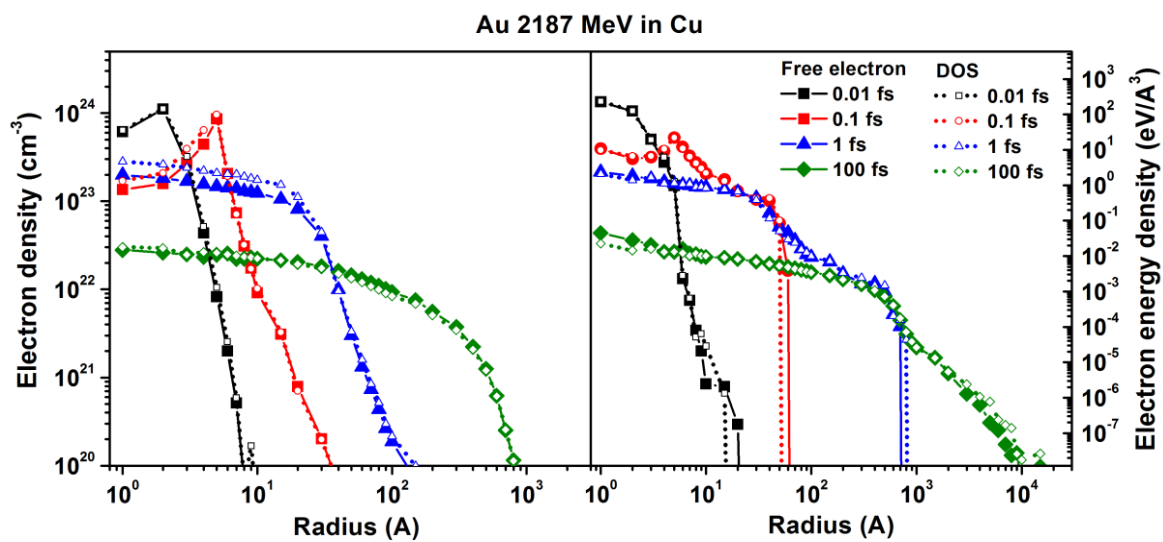

**Figure 9** The density of excited electrons and their energy at different times after passage of Au 2187 MeV ion in Cu calculated within the free-electron and effective one-band dispersion laws.

Figure 9 demonstrates the density of excited electrons and their energy at different times after a passage of Au 2187 MeV ion in Cu. The results of two different simulations are compared: within effective one-band approximation vs. without any corrections. It is clear from Figure 9 that the correction to the IMFP given by DOS effective dispersion relation is not very significant for spatial spreading of electrons as well as the excess energy contained in the electron subsystem. Although a choice of the model cross sections affects the calculated number of low-energy electrons generated in an SHI track, their contribution into the spatial redistribution of the electronic excitation is minor at the considered timescales.

## 3. Interaction with a lattice

We analyze two different models of the elastic scattering of electrons: scattering on optical phonons present in CDF from optical experiments vs. nonrelativistic Mott's atomic cross sections of elastic scattering of a free electron on individual target atoms [62,63].

For materials where optical phonons exist, the complex dielectric function reconstructed from the experimental optical data contains low-energy (<0.1 eV) peaks that can be associated with excitation of optical phonons [33] (fitted CDFs coefficients including optical phonon contributions for considered solids can be found in [1]). Determining the cross section from Eq.(1), we evaluate the mean free paths of electron interaction taking $E_{0i}(q) = E_{0i} + \hbar^2 q^2 / (2M)$ with $M$ equal to the average mass of atom of the target as the energy-momentum dependence of oscillators related to optical phonons in Eq.(2). Here we do not apply other possible atom/phonon dispersion curves. This topic is considered for future research.

Comparison of the elastic electron mean free paths for olivine and LiF calculated using these cross section models (CDF vs Mott's atomic) are given in Figure 10 and Figure 11. As it can be seen, the elastic mean free paths differ drastically between these two models. Therefore, experimental validation of the models is absolutely necessary.

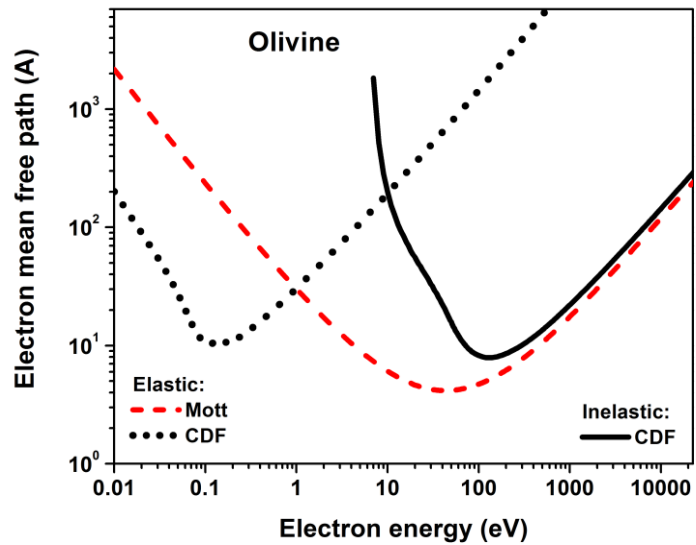

**Figure 10** The mean free paths of electrons in Olivine.

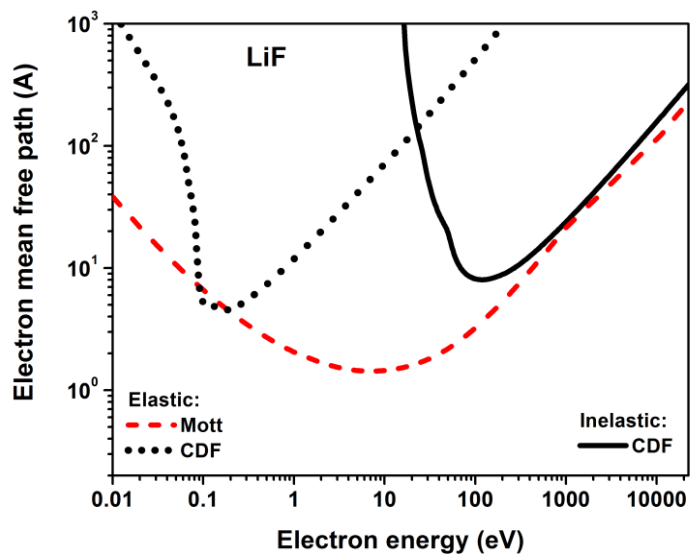

**Figure 11** The mean free paths of electrons in LiF.

Further we analyze how these cross sections affect electron spreading, in particular, how the ballistic transport turns into a diffusive one. Figure 12 demonstrates the temporal dependence of the angular distributions of generated electrons at times up to 100 fs after passage of Au 2187 MeV in $Al_2O_3$. It is clearly seen from the figure that the largest part of the primary electrons (at 0.01 and 0.1 fs) moves perpendicularly to the ion trajectory.

The distribution considerably changes already after 1 fs tending to the uniform one only after 10 fs. This confirms diffusive propagation of exited electrons at times larger than ten femtoseconds after the ion passage. Only after this time, spreading of electrons and their excess energy approaches diffusive behavior (except for its very front), which is assumed e.g. in the framework of the thermal spike model [57].

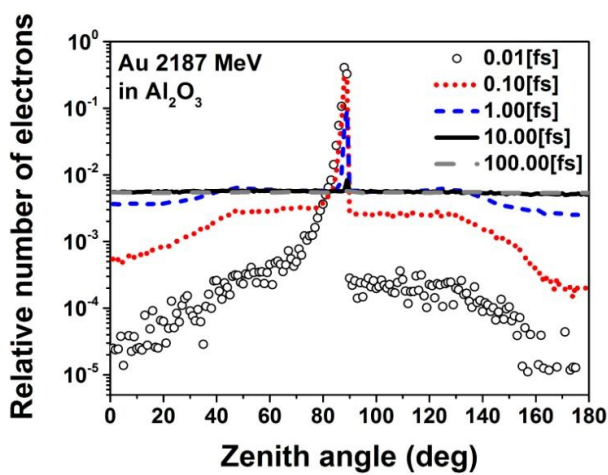

**Figure 12** The temporal dependence of the zenith (θ) angular distribution of electrons after a passage of Au 2187 MeV ion in $Al_2O_3$. The distribution is calculated using the elastic cross sections from the CDF and normalized to the total number of ionized electrons.

Simulations including and excluding scatterings of electrons on the target lattice are compared in order to study which of the scattering processes (elastic vs inelastic) is more important for randomization of the angular distribution. The results for Au 2187 MeV ion in $Al_2O_3$ in Figure 13 demonstrate that the uniform angular distribution of the ionized electrons is reached much faster when elastic interactions of fast electrons with target atoms (optical phonons) are included.

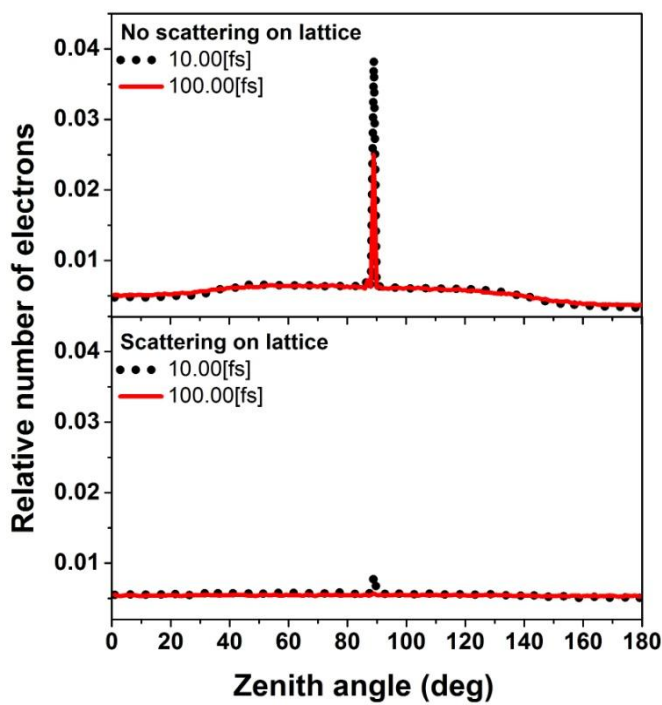

**Figure 13** The temporal dependence of the angular distribution of electrons after impact of Au 2187 MeV in $Al_2O_3$ evaluated with and without scatterings of electrons on the lattice (elastic scattering on optical phonons) in an ion track at 10 and 100 fs. The distributions are normalized to the total number of ionized electrons.

The radial distributions of the energy density and the concentration of generated electrons at 10 and 100 fs after a passage of Au 2187 MeV in LiF are presented in Figure 14. Different approximations for scattering of fast electrons on the target lattice are compared: (i) no such interaction is taken into account, (ii) CDF-based cross sections are used in simulations, (iii) simulations with Mott's atomic cross sections [35]. The figure shows that the difference in the radial distributions of the fastest fraction of electrons (the first front, traveled distances <1000-5000 Å at 10-100 fs) is only minor. The slower part of electrons is affected much stronger; the different kinds of cross sections change the position and shape of the second front of propagation of electronic excitations. Therefore, accurate description of elastic scattering of electrons is important for modeling of electron kinetics in a track. Unfortunately, there is a lack of experimental data to verify the modeling results.

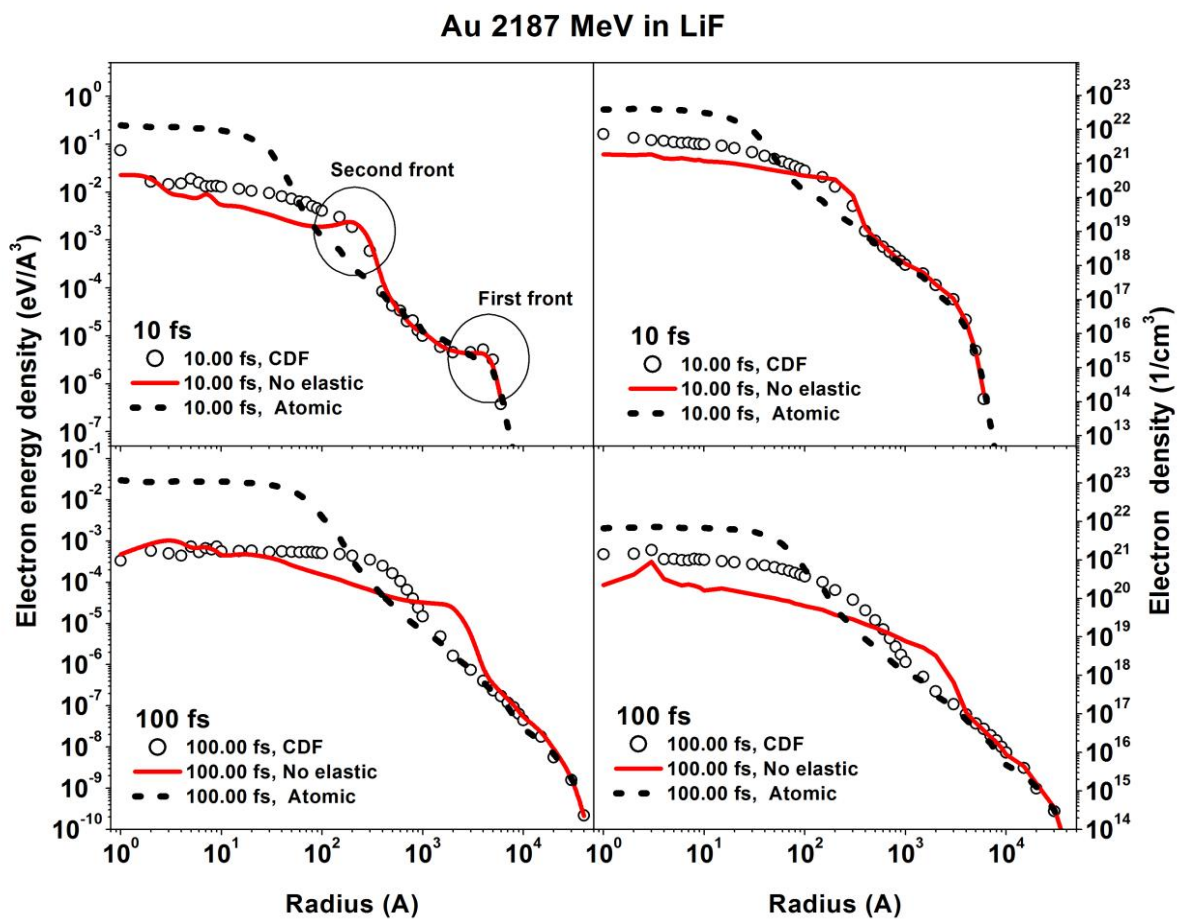

**Figure 14** The radial distributions of electrons and their energy calculated without and with elastic scattering on lattice in the frames of the CDF formalism and the atomic approximation [11].

Because the effect of scattering of valence holes on a lattice in SHI tracks was detected in experiments [32,64], we included in the extended TREKIS code motion of valence holes in dielectrics analogously to that of excited electrons [36]. The dispersion relation of a hole was determined from the density of states of the material within the effective one-band approximation [26]. As discussed in the Ref. [1], in the present model we do not introduce any interaction among excited free electrons and valence holes, neither within the holes subsystem itself. It may lead to transient charge non-neutrality. As it was shown in [32], mobility of holes changes the radial distribution of the dose at longer (picoseconds) times. Considering the mass differences between a valence hale and a target atom/ion, any possible interaction between holes and electrons would lead rather to electrons and holes ambipolar diffusion, than to a Coulomb explosion in ionic system.

Figure 15 shows the densities of generated electrons and valence holes after a passage of an Au ion in silicon. In contrast to LiF, valence holes can impact-ionize valence band of Si generating new electrons, because valence holes can have enough energy to bring an electron across the band gap ($E_{gap} < E_{VB}$). The figure demonstrates that the density of exited electrons increases significantly at longer times (~100 fs) due to impact ionizations by valence holes (Figure 15, right panel).

The left panel of Figure 15 shows that valence holes in silicon propagate fast from the track center, which is the result of their small effective mass and large mean free paths. Consequently, in the case of Si, valence holes bring a considerable part of the deposited energy away from the track core. As was discussed in Ref. [1], a fast electron escape from the track core precludes Si from nonthermal melting in SHI tracks; here we confirm that valence holes also propagate from the track core sufficiently fast preventing crystalline silicon from any nonthermal melting or phase transitions in SHI tracks.

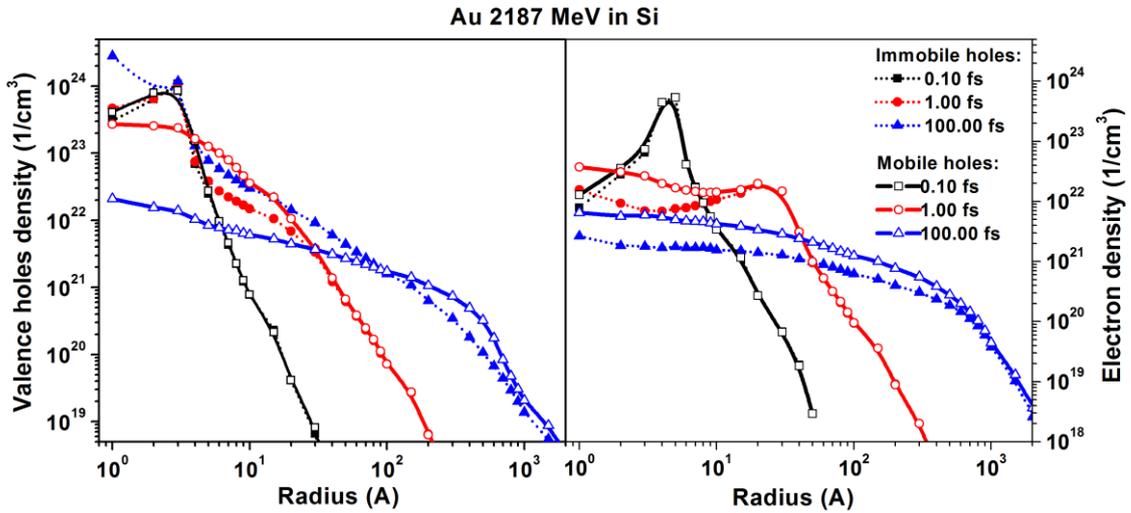

**Figure 15** The radial densities of valence holes (left panel) and ionized electrons (right panel) at different times after a passage of Au 2187 MeV ion in silicon.

## 4. Photon transport

Figure 16 demonstrates radial distributions of the density of photons as well as of their energy, generated due to radiative decay of deep shell holes in a track of Au 2187 MeV in LiF. The initial total energy accumulated in these photons reaches ~4% of the total energy deposited by an SHI and then drops to zero at tens of fs due to photabsorption by target electrons, predominantly from the valence band.

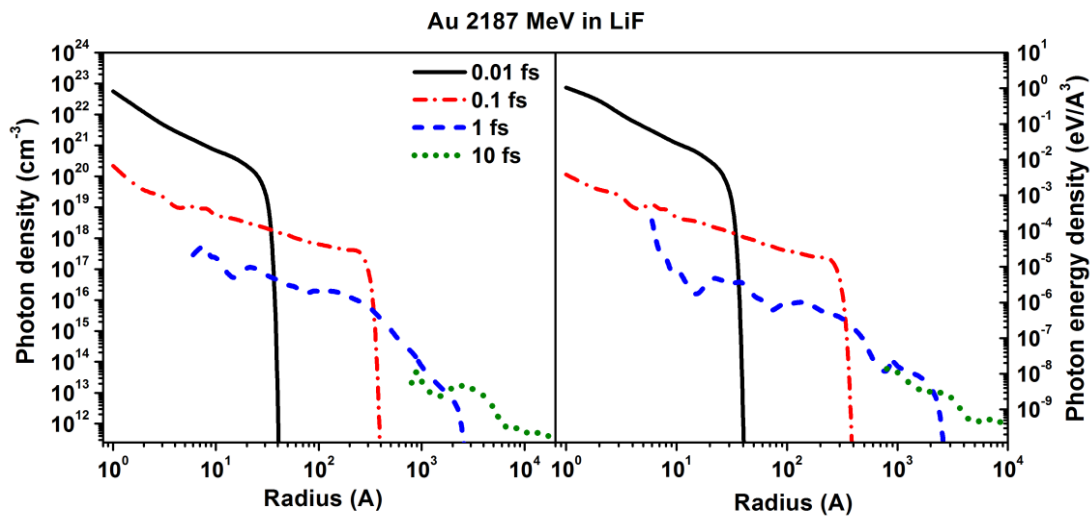

**Figure 16** The radial densities of photons and their energy at different times after a passage of Au 2187 MeV ion in LiF.

Figure 17 demonstrates a comparison of the densities of electrons and the energy they accumulate at different times calculated taking into account radiative decay of core holes vs without this process. It can be clearly seen that travelling with the speed of light photons reach farther distances from the ion trajectory much faster than delta-electrons and bring a part of the energy outward from the track core, ionizing new electrons at longer distances. However, the density of these additionally photo-ionized electrons is small.

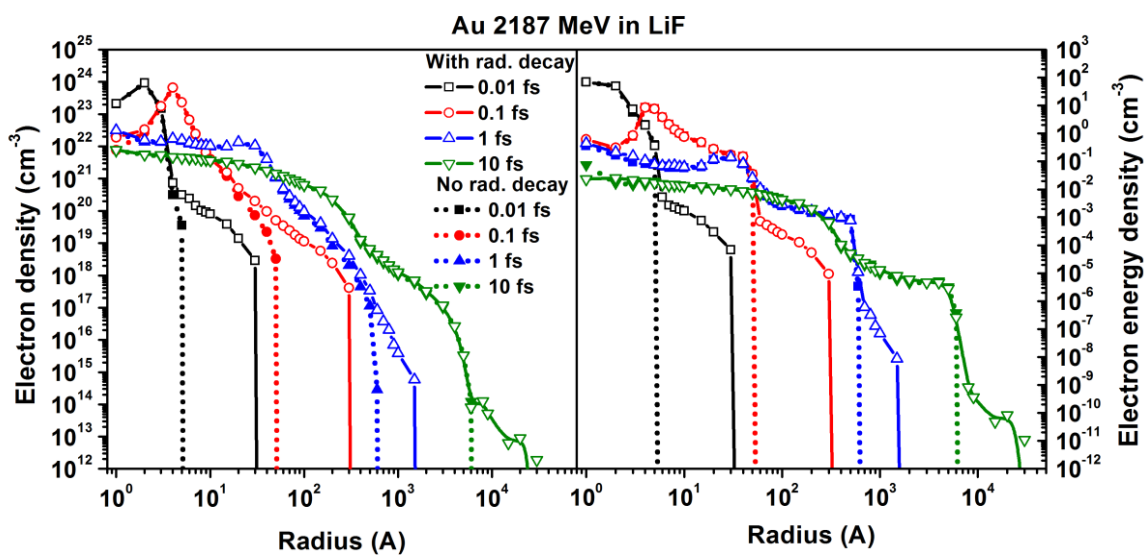

**Figure 17** The radial distributions of electrons calculated without radiative decays of deep holes and with these decays and induced photon transport at different times after passage of Au 2187 MeV ion in LiF.

## Conclusions

In this paper we analyzed effects of different model forms of cross sections used in Monte-Carlo simulations (on example of TREKIS code [1]) of the electronic kinetics in SHI tracks in solids. In particular:

Effects of different energy-momentum dependencies of the artificial oscillators used in the CDF are investigated. It is shown that the free-electron approximation works well for high energies of electrons, whereas for low energy electron impacts, effects of band structure of the material cannot be neglected. Application of the 'effective one-band' approximation significantly improves the electron mean free path, reproducing experimental data very well in the case of metal targets. For semiconductors (e.g. Si), an agreement between the experiments and calculations is not as good for electron energies below ~50 eV.

For elastic scattering cross sections, demonstrating an effect of collective lattice response, two different models were analysed (based on CDF and atomic approximations). Large differences in the electron-lattice energy exchange rates and randomization of electron motion were detected. Therefore, further studies and comparisons with experiments are required to justify a choice of a model.

It was also demonstrated that elastic scattering is the dominant channel of randomization of the directions of moving electrons. Excluding elastic scattering provides much longer times for ballistic transport of electrons.

It was shown that the kinetics of valence holes significantly affects redistribution of the excess electronic energy in the vicinity of the SHI trajectory as well as its conversion into lattice excitation in dielectrics and semiconductors [36].

We showed that photon transport plays only a minor role in the electron kinetics because core holes are predominantly decaying via Auger-decay channel. Although the relative energy they accumulate is very small for the considered range of SHI parameters, emitted photons can travel large distances before their absorption.

## ACNOWLEGEMENTS

R.A. Rymzhanov and A.E. Volkov thank a support from Russian Foundation for Basic Research (grant 16-32-50053). Partial financial support from the Czech Ministry of Education (Grants LG15013 and LM2015083) is acknowledged by N.A. Medvedev. A.E. Volkov also acknowledges financial support from grants 15-58-15002 and 15-02-02875 of Russian Foundation for Basic Research, as well as from Project 13 HALO of Russian Ministry of Education, and from the Ministry of Education and